\documentstyle{elsart}

\begin{document}

 \begin{frontmatter}
   \title{Casimir Force between two Half Spaces of Vortex Matter in
   Anisotropic Superconductors} 
   
   \author[zuerich]{H.~P.\ B\"uchler},
   \author[zuerich,santacruz]{H.~G.\ Katzgraber},
   \author[zuerich]{and G.\ Blatter}
   \address[zuerich]{Theoretische Physik, Eidgen\"ossiche Technische
     Hochschule-H\"onggerberg, CH-8093 Z\"urich, Switzerland}
   \address[santacruz]{
     Department of Physics, University of California, Santa Cruz,
     California 95064 }
   
   \begin{abstract}
    We present a new approach to calculate the attractive long-range
    vortex-vortex interaction of the van der Waals type present in anisotropic 
    and layered superconductors. The mapping of the statistical mechanics of
    two-dimensional charged bosons allows us to define a Casimir problem: Two
    half spaces of vortex matter separated by a gap of width $R$ are mapped to 
    two dielectric half planes of charged bosons interacting via a massive
    gauge field. We determine the attractive Casimir force between the two
    half planes and show that it agrees with the pairwise summation of the van
    der Waals force between vortices.
   \end{abstract}
   
 \end{frontmatter}

 \section{Introduction}

\begin{figure}[ht]
   \makebox[12cm][c]{
     \psfig{figure=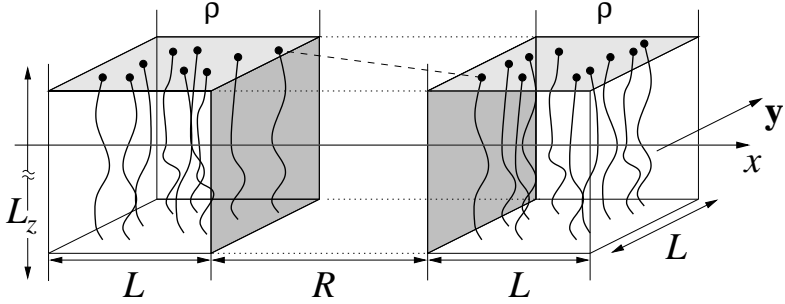,height=4.02cm,width=10.34cm}}
   \caption{Geometry used for the calculation of the Casimir force. The $ab$
     plane of superconductor is parallel to the $xy$ plane, while the $c$ axis 
     is along the applied magnetic field in $z$ direction. The distance
     between the two half spaces is $R$. }  \label{fig}
 \end{figure}

 Vortices in isotropic type-II superconductors repel each other with
 the interaction strength decaying exponentially on the scale  $R  > \lambda$
 ($R$ is the distance between the vortices, while $\lambda$ denotes the
 London penetration length). However, it
 has recently been shown \cite{blatter96,katzgraber99} that in layered
 superconductors 
 the thermal fluctuations of the flux lines give rise to a long-range
 {\em attraction} $V_{{\rm vdW}} \sim - (T/d)(\lambda/R)^{4}$ of the van der
 Waals type on the scale
 $R > \lambda$, where $T$ denotes the temperature and $d$ is the interlayer
 distance. The equivalence of three dimensional statistical mechanics with the
 $2+1$ dimensional imaginary time quantum mechanics  allows us to describe
 the flux lines as two dimensional charged bosons with an interaction mediated 
 by a gauge field ${\bf a}$. In the case of vortex matter 
 the material properties are mapped to a dielectric
 permittivity for the gauge field. Geometric boundary conditions and
 dielectric permittivities then
 cause a shift in the zero-point energy of the gauge field
 known as the Casimir effect. It is subject of the present work to calculate
 the Casimir force between two half spaces of vortex matter separated by a
 vortex free region as shown in Fig.~\ref{fig}.
 It is well known \cite{lifshitz56} that in special geometries (e.~g.\ two
 dielectric half spaces) van der Waals
 forces can be related to the Casimir force  via  pairwise summation. This
 interpretation of the Casimir force allows us to derive the van der Waals
 force between flux lines in 
 anisotropic superconductors via the Casimir approach.  
 \input{psfig}

 \section{Casimir Force between two half Spaces of Vortex Matter} 
 
 {\em Vortices as 2D Bosons}: Within the London theory the free energy of vortices in an isotropic
 superconductor takes the form 
 \begin{equation}
   {\mathcal F} = \frac{\epsilon_{0}}{2} \int d^{3}{\bf x} \:  d^{3}{\bf y} \:
   {\bf j}({\bf x}) 
   \: 
   \frac{e^{-  
       \lambda |{\bf x}-{\bf y}|}}{|{\bf x}-{\bf y}|} \:  {\bf j}({\bf y}) 
 \end{equation}
 with the current ${\bf j}= ({\bf J}, \rho) = \sum_{\mu} (\partial_{z} {\bf R} 
 , 1) \delta^{2}\left({\bf R} - {\bf R}_{\mu} \right)$ describing the
 vortex lines. Following a suggestion of Nelson \cite{nelson}, the statistical 
 mechanics of vortices can be mapped to the imaginary time quantum mechanics
 of two-dimensional (2D) bosons. The $c$ axis of the superconductor is  mapped 
 to the imaginary time of the bosons $z \rightarrow \tau$, the temperature $T$
 becomes the Planck constant $\hbar^{B}$, while
 the interaction between the vortices is
 mediated by a fake gauge field ${\bf a}$ with a coupling $g^{2} = 4 \pi
 \varepsilon_{0}$, where $\epsilon_{0}=(\Phi_{0}/4 \pi \lambda)^{2}$ is the
 line energy \cite{popov}. The action of the 2D bosons becomes
 \begin{equation}
   \begin{array}{l}
     {\displaystyle {\mathcal S}\left[ {\bf j}, {\bf a} \right] = \int d\tau
       \sum_{\mu} \left\{ 
         \frac{m}{2} \left[ \partial_{\tau} {\bf R}_{\mu}(\tau)
         \right]^{2} - 
         \mu^{B} \right\}  + \int d\tau
         d^{2}R {\Bigg \{ } i \:  {\bf a} \cdot {\bf j} } \\
       \hspace{80pt} 
           {\displaystyle \left. + 
         \frac{1}{2 g^{2} \lambda^{2}} {\bf a}^{2} +
      \frac{1}{2 g^{2}} \left( \nabla
           \times {\bf a} 
           \right)^{2}_{xy} + 
           \frac{c^{2}}{2 g^{2}} \left( \nabla \times {\bf a} \right)_{\tau}
         \right\}}  
     \end{array} \label{2dbosons}
 \end{equation}
 with the bare  boson mass $m=\varepsilon_{0}$. The
 self-interaction of the vortices via  the gauge field leads to a mass
 renormalization 
 $m\rightarrow m^{B}= \varepsilon_{l}$ where $\varepsilon_{l}$ is the
 dispersive line tension \cite{blatter94} of the vortex line. The
 anisotropy $\epsilon$ between the $ab$ plane and the
 $c$ axis is introduced 
 by $c=1/\epsilon$ (the light velocity in the boson
 system).   
 
 {\em Material Properties}: We can map the material properties of the vortex matter to a dielectric
 permittivity  $\epsilon_{V}$  by integrating 
 over the currents ${\bf J}$ of the 2D bosons, leading to a term $
 (\rho/m){\bf a}_{xy}^{2}$ in the action. Performing functional
 derivatives leads to the dispersion relation for
 the  gauge field  ${\bf a_{xy}}$
 \begin{equation}
   \left[ \frac{\omega^{2}}{c^{2}} \epsilon_{V}(\omega) + k^{2} +
     \frac{\epsilon^{2}}{\lambda^{2}}\right] {\bf a}_{xy} = 0 \hspace{20pt}
   \mbox{with} \hspace{10pt}
   \epsilon_{V}(\omega) = 1 + \frac{g^{2} \rho}{m^{B} \omega^{2}} \: .
 \end{equation} 
 The boundary conditions at the interface between the vortex matter and the
 vortex free region together with the dispersion relation and the
 dielectric permittivity define a Casimir problem \cite{lifshitz56}.

 {\em Casimir Force}: The Casimir force of the 2D dielectric system becomes
 \begin{equation}
   \begin{array}{c}
     {\displaystyle f= -\frac{\hbar^{B}}{\pi^{2}} \int_{1}^{\infty} dp
       \int_{0}^{\infty} d\omega \frac{s p \omega^{2}}{c^{2} \sqrt{p^{2}-1}}} 
     \hspace{180pt} \\ \hspace{50pt}
     {\displaystyle \times \left[ \left( \frac{\left[ 1+
               (\lambda \omega)^{-2} \right] s_{\epsilon} + \left[ \epsilon_{V}
               +(\lambda 
               \omega)^{-2} \right] s}{\left[ 1+
               (\lambda \omega)^{-2} \right] s_{\epsilon} - \left[ \epsilon_{V}
               +(\lambda  \omega)^{-2} \right] s} \right)^{2}e^{2 R
           \omega/c} -1 \right]^{-1} 
       }\:, 
   \end{array}
 \end{equation} 
 where $p^{2} = 1+c^{2} k^{2} \omega^{-2}$, $s_{\epsilon}^{2} =
 \epsilon_{V} -1 + p^{2} +(\lambda \omega)^{-2}$ and $ s^{2} = p^{2}+ (\lambda
 \omega)^{-2}$. This expression is determined by three frequencies:
 $\omega_{d} = \pi/d$ is an 
 upper cut-off due to the layered structure of the superconductor, while
 $\omega_{R}=1/\epsilon R$ derives from the geometric length, and 
 $\omega_{\lambda} = 1/\lambda$ describes the mass of the gauge field. 
 
 {\em Pairwise Summation}: In the following we present the
 Casimir force for the three different length scales. The pairwise summation
 (i.~e.\  the summation of the van der Waals forces 
 between the vortex lines in each half space) then provides the van der Waals
 force  between flux lines. For intermediate
 distances $R< d \epsilon^{-1},\lambda \epsilon^{-1}$ the cut-off $\omega_{d}$
 is relevant and we 
 obtain the result
 \begin{equation}
   \begin{array}{rcl} 
     {\displaystyle f= - \frac{\left(\epsilon_{V} -1 \right)^{2}}{16 \pi^{2}}
       \frac{\hbar^{B}  \omega_{d}}{R^{2}} } &
     \hspace{10pt} \Longrightarrow \hspace{10pt} & \displaystyle{ V_{{\rm
           vdW}} = 
       - \frac{4 
         \epsilon_{0}}{\ln^{2}(\pi \lambda 
         d^{-1})} \frac{T}{d \epsilon_{0}}\left(\frac{\lambda}{R}\right)^{4}}
   \end{array} \: .
 \end{equation}
 At larger distances $d \epsilon^{-1} < R < \lambda \epsilon^{-1}$ retardation
 of the gauge field becomes important and the frequency $\omega_{d}$ is
 replaced by $ c/R$
 \begin{equation}
   \begin{array}{rcl}
     {\displaystyle f= - \frac{19 \left(\epsilon_{V} -1 \right)^{2}}{1024 \pi}
       \frac{\hbar^{B} 
         c}{R^{3}} }& \hspace{10pt} \Longrightarrow \hspace{10pt}
      & {\displaystyle V_{{\rm vdW}}^{{\rm ret}} = - \frac{(171
         \pi  / 256) 
         \epsilon_{0}}{\ln^{2}(\pi \lambda 
         (\epsilon R)^{-1})} \frac{T}{\lambda \epsilon 
         \epsilon_{0}}\left(\frac{\lambda}{R}\right)^{5} }
   \end{array} \: .
 \end{equation} 
 At very large distances the mass of the gauge field leads to an exponential
 decay 
 \begin{equation}
  \begin{array}{rcl}
  {\displaystyle  f = - \frac{\hbar^{B} c}{R\: \lambda^{2}} \: 
    \frac{8 \pi \lambda^{4} \rho^{2}}{ c^{2}} e^{- \frac{R}{\lambda c}}}&
  \hspace{10pt} \Longrightarrow \hspace{10pt}   & 
   {\displaystyle  V_{{\rm vdW}} = - 4 \sqrt{\pi} \varepsilon_{0} \frac{T
       \epsilon^{4}}{\varepsilon_{0} \lambda} \left(\frac{\lambda}{\epsilon
         R}\right)^{\frac{3}{2}} e^{- 2 \frac{\epsilon 
       R}{\lambda}}} 
 \end{array} \: .
 \end{equation}
 The van der Waals forces at intermediate and large distances agree with the
 derivation by Blatter and Geshkenbein \cite{blatter96}.

 \input{psfig}
 \begin{figure}[ht]
   \makebox[12cm][c]{
     \psfig{figure=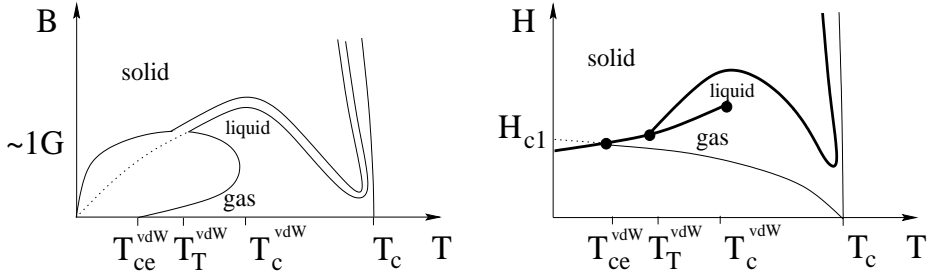,height=3.56cm,width=12.3cm}}
   \caption{Phase diagrams: $T_{ce}^{\rm vdW}$ denotes  a critical endpoint,
     $T_{ce}^{\rm vdW}$ is a 
 critical point, and $T_{T}^{\rm vdW}$ is a triple  point  }  \label{fig2}
 \end{figure}
 {\em Phase Diagram}: The attraction
 between flux lines leads to interesting modifications of the $B-T$ phase
 diagram of anisotropic type-II superconductors, see Fig.~\ref{fig2} for a
 schematic drawing.  At low
 temperature $T<T_{ce}^{\rm vdW}$ a first order phase transition takes  the
 Meissner state into the vortex solid ( with the critical field $H_{c_{1}}$
 lowered by the attraction), while at higher
 temperature $T_{c} > T >T_{ce}^{\rm vdW}$ the Meissner state goes into a
 vortex gas through a second order phase transition. In the temperature 
 range $T_{T}^{\rm vdW}< T <T_{c}^{\rm vdW}$ we can distinguish  a low
 density vortex gas from a high density vortex liquid. Concurrent to the first 
 order jump in the magnetization  a phase separation is observed and the 
 domains of vortex matter separated by vortex free regions interact via the
 Casimir force described in this paper.

\end{document}